# Spinning superconducting electrovacuum soliton

Irina Dymnikova

*Department of Mathematics and Computer Science, University of Warmia and Mazury,
Żołnierska 14, 10-561 Olsztyn, Poland; e-mail: irina@matman.uwm.edu.pl*

In nonlinear electrodynamics coupled to general relativity and satisfying the weak energy condition, a spherically symmetric electrically charged electrovacuum soliton has obligatory de Sitter center in which the electric field vanishes while the energy density of electromagnetic vacuum achieves its maximal value. De Sitter vacuum supplies a particle with the finite positive electromagnetic mass related to breaking of space-time symmetry from the de Sitter group in the origin. By the Gürses-Gürsey algorithm based on the Newman-Trautman technique it is transformed into a spinning electrovacuum soliton asymptotically Kerr-Newman for a distant observer. De Sitter center becomes de Sitter equatorial disk which has both perfect conductor and ideal diamagnetic properties. The interior de Sitter vacuum disk displays superconducting behavior within a single spinning soliton. This behavior found for an arbitrary nonlinear lagrangian $\mathcal{L}(F)$, is generic for the class of regular spinning electrovacuum solutions describing both black holes and particle-like structures.

**Introduction** - The Kerr-Newman geometry [1]

$$ds^2 = -dt^2 + \frac{\Sigma}{\Delta}dr^2 + \Sigma d\theta^2 + \frac{(2mr - e^2)}{\Sigma}(dt - a\sin^2\theta d\phi)^2$$

$$+ (r^2 + a^2)\sin^2\theta d\phi^2; \quad A_i = -\frac{er}{\Sigma}[1; 0, 0, -a\sin^2\theta] \quad (1)$$

where $A_i$ is associated electromagnetic potential, and

$$\Sigma = r^2 + a^2\cos^2\theta; \quad \Delta = r^2 - 2mr + a^2 + e^2, \quad (2)$$

was derived by the complex coordinate transformation from the Reissner-Nordström metric with using an algebraic trick discovered by Newman and Janis [2].

Carter has found that the parameter $a$ couples with the mass $m$ and charge $e$ independently, giving the angular momentum $J = ma$, an asymptotic magnetic momentum $\mu = ea$, and the same gyromagnetic ratio as predicted for a spinning particle by the Dirac equation [3].

The Kerr-Newman solution belongs to the Kerr family of solutions to the source-free Maxwell-Einstein equations, and represents the exterior fields of a rotating charged body [3]. The question of an interior material source for these exterior fields has been addressed in a lot of papers. The source models for the Kerr-Newman interior can be roughly divided into disk-like [4–6], shell-like [7–9], bag-like [10–15], and string-like ( [16] and references therein).

In the Kerr-Newman geometry the surfaces $r = const$ are the oblate ellipsoids

$$r^4 - (x^2 + y^2 + z^2 - a^2)r^2 - a^2 z^2 = 0 \quad (3)$$

which degenerate, for $r = 0$, to the equatorial disk

$$x^2 + a^2 \leq a^2, \quad z = 0 \quad (4)$$

centered on the symmetry axis and bounded by the ring

$$x^2 + y^2 = a^2, \quad z = 0 \quad (5)$$

which comprises the Kerr ring singularity [17].

The coordinates $x, y, z$ are related to the Boyer-Lindquist coordinates $r, \theta, \phi$ by

$$x^2 + y^2 = (r^2 + a^2)\sin^2\theta; \quad z = r\cos\theta \quad (6)$$

The main disaster of the Kerr-Newman geometry discovered by Carter is nontrivial causality violation in the case of a charged particle, $a^2 + e^2 > m^2$ [3]. In this case there are no Killing horizons and the manifold is geodesically complete (except for geodesics which reach the singularity), and any point can be connected to any other point by both a future and a past directed timelike curve. The condition of the causality violation [3]

$$r^2 + a^2 + \Sigma^{-1}(2mr - e^2)a^2 \sin^2\theta < 0 \quad (7)$$

is satisfied in the vicinity of the Kerr disk where the vector $\partial/\partial\phi$ is timelike, but closed timelike curves entering the region (7) can extend over the whole space and cannot be removed by taking a covering space [3].

Two ways of exploring the Kerr-Newman geometry were marked by Israel [4] who noticed that $r$ tends to zero at approaching the disk, but $\nabla r$ does not vanish. Due to intrinsic meaning of the coordinate $r$, this leads to the two-sheeted geometry with closed timelike curves even in the case $e = 0$ or a discontinuity of the metric (defined on the manifold with $r \geq 0$) across the disk [4].

Two lines of research were thus to remove the "two-sheetedness" of the Kerr-Newman geometry by truncation of the negative-$r$ sheet either to give it physical interpretation. The second approach resulted in the various singular stringy structures ( [16] and references therein).

Carter found that the gravitational field becomes repulsive in the vicinity of the disk [3]: all timelike and null geodesics avoid the ring singularity (except for some in the equatorial plane), for a particle falling down the axis, the gravity becomes repulsive when $r < |a|$ [18].

Assuming that the Kerr-Newman metric is the field of a massive charged layer over the equatorial disk (4), Israel applied the well understood theory of surface layers in general relativity and revealed the basic feature of the Kerr-Newman geometry: appearance of a negative pressure as a source of the gravitational repulsion [4].

Studying electromagnetic fields of a rotating body which would give the Kerr-Newman behavior outside, Tiomno found that one can expect in the interior region a perfect conductor behavior [12].

Gürses and Gürsey proposed an approach based on the Trautman-Newman technique (which includes the Newman-Janis trick), and derived the Kerr-Schild metrics [19] in general form including a model-independent description of an interior for a spinning systems [20].

In this approach the metric in the Kerr-Schild form

$$g_{\mu\nu} = \eta_{\mu\nu} + \frac{2f(r)}{\Sigma} k_\mu k_\nu \quad (8)$$

where $\eta_{\mu\nu}$ is the Minkowski metric and $k_\mu$ is a vector null field tangent to the Kerr principal null congruence, involves a function $f(r)$ which comes from a spherically symmetric solution [20]. The coordinate $r$ is defined as an affine parameter along either of two principal null congruences, and the surfaces of constant $r$ are confocal ellipsoids (3). For the Kerr-Newman geometry

$$f(r) = mr - e^2/2 \quad (9)$$

In the Boyer-Lindquist coordinates the metric is

$$ds^2 = \frac{2f - \Sigma}{\Sigma} dt^2 + \frac{\Sigma}{\Delta} dr^2 + \Sigma d\theta^2 - \frac{4af \sin^2\theta}{\Sigma} dt d\phi +$$

$$\left( r^2 + a^2 + \frac{2fa^2 \sin^2\theta}{\Sigma} \right) \sin^2\theta d\phi^2 \quad (10)$$

where now

$$\Delta = r^2 + a^2 - 2f(r) \quad (11)$$

The anisotropic stress-energy tensor responsible for (10) [20] can be written as

$$T_{\mu\nu} = (\rho + p_\perp)(u_\mu u_\nu - l_\mu l_\nu) + p_\perp g_{\mu\nu} \quad (12)$$

in the orthonormal tetrad with the time-like vector $u^\mu$ and three space-like vectors:

$$u^\mu = \frac{1}{\sqrt{\Delta\Sigma}}[(r^2+a^2)\,\delta_0^\mu + a\,\delta_3^\mu], \quad l^\mu = \sqrt{\frac{\Delta}{\Sigma}}\,\delta_1^\mu,$$

$$n^\mu = \frac{1}{\sqrt{\Sigma}}\delta_2^\mu, \quad m^\mu = -\frac{1}{\sqrt{\Sigma}\sin\theta}[a\sin^2\theta\,\delta_0^\mu + \delta_3^\mu] \quad (13)$$

The eigenvalues of the stress-energy tensor (12), its components in the co-rotating references frame where each of ellipsoidal layers of constant $r$ rotates with the angular velocity $\omega(r) = u^\phi/u^t = a/(r^2+a^2)$ [15], are given by

$$T_{\mu\nu} u^\mu u^\nu = \rho(r,\theta); \quad T_{\mu\nu} l^\mu l^\nu = p_r = -\rho;$$

$$T_{\mu\nu} n^\mu n^\nu = T_{\mu\nu} m^\mu m^\nu = p_\perp(r,\theta) \quad (14)$$

They are related to the function $f(r)$ by

$$\kappa\Sigma^2 \rho = 2(f'r - f); \quad \kappa\Sigma^2 p_\perp = 2(f'r - f) - f''\Sigma \quad (15)$$

where $\kappa = 8\pi G$. All the metrics from the Kerr-Schild class are also solutions of the Gürses-Gürsey equations with the source term (12) [20].

Burinskii has noted that appearance of the negative pressure in the KN interior models [4,6,9,12,13,21], suggests superconducting properties of a source material [14,22–25].

Looking for regular sources with a negative pressure, Elizalde and Hildebrandt [26] adopted generalization of the energy-matter content called spherically symmetric anisotropic vacuum [27,28] specified by

$$T_0^0 = T_1^1 \quad (16)$$

for a non-rotating observer, which implies $p_r = -\rho$.

Burinski, Elizalde, Hildebrandt and Magli applied then the Kerr-Schild metric in the Gürses-Gürsey form for regular (anti) de Sitter sources matched to the Kerr-Newman exterior at the domain wall boundary. The Kerr singular ring is regularized by an anisotropic rotating source, so that the negative-$r$ sheet is absent [15].

Among the models with the postulated regular de Sitter core matched directly to the Kerr-Newman metric ( [14,15] and references therein), there is the model involving nonlinear electrodynamics (NED) [29]. The electrically charged external KN metric is matched to the magnetically charged internal core at the point $r_m$ where the electric susceptibility diverges, so that the sphere $r = r_m$ looks like an ideal conducting surface which confines the magnetic charge. This type vacuum core expels magnetic charges by construction, so that one can speculate [29] that it possesses superconducting properties.

The problem of matching the Kerr-Newman exterior to a rotating material source does not have a unique solution, since one is free to choose arbitrarily the boundary between the exterior and the interior [4], not speaking about freedom in choosing an interior model itself.

The aim of this Letter is to show that in nonlinear electrodynamics coupled to gravity the field equations admit regular solutions describing a spinning electromagnetic soliton (a regular finite-energy solution of the nonlinear field equations, localized in the confined region and holding itself together by its own self-interaction), and to reveal its basic model-independent features.

**Spinning electrovacuum soliton** - A vacuum with the reduced symmetry (16) as compared with the maximally symmetric de Sitter vacuum (all eigenvalues $T_i^i$ equal and constant), has the advantage that reducing symmetry allows vacuum density to become evolving and clustering [28]. Spherically symmetric space-time specified by (16), has obligatory de Sitter center and mass of an object related to de Sitter vacuum trapped inside and smooth breaking of space-time symmetry from the de Sitter group in the origin [30]. Stress-energy tensor of an electromagnetic field belongs to the class (16).

The hybrid model [29] was obtained by modification of the Ayön-Beato-Garcia solution [31], as a possible circumvention of the Bronnikov theorem on the non-existence of regular electrically charged structures in non-linear electrodynamics [32].

The Bronnikov theorem tells about the regular electric structures satisfying the Maxwell weak filed limit $\mathcal{L} \sim F$ as $r \to 0$, and thus does not embrace all possibilities.

Discarding the requirement of the Maxwell weak field limit as $r \to 0$ (a field must not be weak to be regular), one obtains in nonlinear electrodynamics coupled to gravity and satisfying the weak energy condition, regular electrically charged spherically symmetric solutions with the obligatory de Sitter center in which field tension goes to zero, while the energy density of the electromagnetic vacuum $T_0^0$ achieves its maximal finite value which represents the de Sitter cutoff on the self-energy density [33].

Spherically symmetric solutions satisfying the condition (16) belong to the Kerr-Schild class [15,26]. By the Gürses-Gürsey algorithm they can be transformed into regular solutions for a spinning charged particle.

The basic features of a spinning electrovacuum soliton follow then from the generic features of an electrovacuum spherically symmetric soliton without specifying the particular form of the NED lagrangian $\mathcal{L}(F)$.

In this case the function $f(r)$ in (10) is given by

$$f(r) = r\mathcal{M}(r); \quad \mathcal{M}(r) = 4\pi \int_0^r \tilde{\rho}(x) x^2 dx \quad (17)$$

where $\tilde{\rho}(r) = \tilde{T}_0^0(r)$ is the energy density of a nonlinear spherically symmetric electromagnetic field. For NED solutions satisfying the weak energy condition, $\mathcal{M}(r)$ is everywhere positive function growing monotonically from $\mathcal{M}(r) = 4\pi\rho(0)r^3/3$ as $r \to 0$ to $m - e^2/2r$ as $r \to \infty$. The mass parameter $m$ appearing in a spinning solution, is the finite positive electromagnetic mass [33].

The condition of the causality violation (7) becomes

$$r^2 + a^2 + \Sigma^{-1} 2f(r) a^2 \sin^2\theta < 0 \quad (18)$$

and is never satisfied due to non-negativity of the function $f(r)$. The vector $\partial/\partial\phi$ is spacelike throughout the whole manifold, so that no causality violation occurs.

For $r \to 0$ the function $f(r)$ in (10) approaches de Sitter asymptotic

$$2f(r) = \frac{r^4}{r_0^2}; \quad r_0^2 = \frac{3}{\kappa\rho(0)} = \frac{3}{\Lambda} \quad (19)$$

Taking into account $\Sigma = (r^4 + a^2 z^2) r^{-2}$, we get for the rotating de Sitter vacuum

$$\frac{2f(r)}{\Sigma} = \frac{r^4}{r_0^2} \frac{r^2}{(r^4 + a^2 z^2)} \quad (20)$$

In the equatorial plane it reduces to $\Lambda r^2/3$, so that the Kerr disk $r = 0$ is totally (together with the ring) intrinsically flat. But $\Lambda$ is non-zero throughout the disk.

The non-rotating metrics are asymptotically de Sitter as $r \to 0$ which is asymptotically flat at $r = 0$ with non-zero $\Lambda$. Rotation transforms the point $r = 0$ into the disk (4). The asymptotic (20) represents the rotating de Sitter vacuum with $\Lambda$ spread over the equatorial disk.

Indeed, the eigenvalues of a stress-energy tensor (12) responsible for the metric (10), are

$$\rho = \frac{r^4}{\Sigma^2} \tilde{\rho}(r); \quad p_\perp = \left(\frac{r^4}{\Sigma^2} - \frac{2r^2}{\Sigma}\right) \tilde{\rho}(r) - \frac{r^3}{2\Sigma} \tilde{\rho}'(r) \quad (25)$$

where $\tilde{\rho}(r)$ is a spherically symmetric density profile. The prime denotes the derivative with respect to $r$.

The equation of state in the co-rotating frame is

$$p_r(r,\theta) = -\rho(r,\theta); \quad p_\perp(r,\theta) = -\rho - \frac{\Sigma}{2r} \frac{\partial \rho(r,\theta)}{\partial r} \quad (26)$$

The second equation in (26) can be written as

$$p_\perp + \rho = 2\left(\frac{r^4}{\Sigma^2} - \frac{r^2}{\Sigma}\right) \tilde{\rho}(r) - \frac{r^3}{2\Sigma} \tilde{\rho}'(r) \quad (27)$$

Taking into account that $r$ obeys (3), we get

$$\frac{r^2}{\Sigma} \to 1; \quad \frac{r^4}{\Sigma^2} \to 1 \quad as \quad z \to 0 \quad (28)$$

identically in the whole equatorial plane including the disk, the ring, and the origin.

As a result we have in the equatorial plane

$$p_\perp + \rho = -\frac{r^3}{2\Sigma} \tilde{\rho}'(r) \quad (29)$$

By (28), the density in the equatorial plane is $\rho(r,\theta) = \tilde{\rho}(r)$. When $r \to 0$, $\tilde{\rho}(r)$ approaches the de Sitter limit $\Lambda/\kappa$, so that on the equatorial disk $\rho(r,\theta) = \Lambda\kappa^{-1}$.

For spherically symmetric solutions regularity requires $r\tilde{\rho}'(r) \to 0$ as $r \to 0$ [33]. With taking into account (28) we find the equation of state on the disk

$$p_\perp = -\rho \quad (30)$$

which represents (with taking into account also $p_r = -\rho$) the rotating de Sitter vacuum in the co-rotating frame.

**Elementary superconductivity** - The basic equations in nonlinear electrodynamics coupled to gravity, are obtained from the Lagrangian

$$S = \frac{1}{16\pi G} \int d^4x \sqrt{-g}[R - \mathcal{L}(F)] \quad (31)$$

where $F = F_{\mu\nu}F^{\mu\nu}$ and $\mathcal{L}(F)$ is an arbitrary function with the Maxwell weak field limit $\mathcal{L}(F) \to F$ for large $r$.

The dynamic equations read

$$\nabla_\mu(\mathcal{L}_F F^{\mu\nu}) = 0 \quad (32)$$

where $\mathcal{L}_F = d\mathcal{L}/dF$, and the Bianchi identities give

$$\nabla_\mu {}^*F^{\mu\nu} = 0 \quad (33)$$

An asterisk denotes the Hodge dual [34]; the antisymmetric unit tensor is chosen in such a way that $\eta_{0123} = \sqrt{-g}$.

Non-zero field components compatible with the axial symmetry, $F_{01}, F_{02}, F_{13}, F_{23}$, are related by

$$F_{31} = a\sin^2\theta F_{10}; \quad aF_{23} = (r^2 + a^2)F_{02} \quad (34)$$

in geometry with the metric (10).

The field invariant $F = F_{\mu\nu}F^{\mu\nu}$ reduces to

$$F = 2\left(\frac{F_{20}^2}{a^2\sin^2\theta} - F_{10}^2\right) \quad (35)$$

In terms of the field vectors defined as

$$\mathbf{E} = \{F_{\beta 0}\}; \mathbf{D} = \{\mathcal{L}_F F^{0\beta}\}; \mathbf{B} = \{{}^*F^{\beta 0}\}; \mathbf{H} = \{\mathcal{L}_F {}^*F_{0\beta}\} \quad (36)$$

the field equations (32)-(33) take the conventional form of the Maxwell equations.

The electric induction $\mathbf{D}$ is connected with the electric field intensity $\mathbf{E}$ by

$$D_\alpha = \epsilon_{\alpha\beta} E_\beta \quad (37)$$

where $\epsilon_{\alpha\beta}$ is the tensor of the dielectric permeability. Symmetry of the oblate ellipsoid (3) gives two independent non-zero eigenvalues

$$\epsilon_r = \frac{(r^2 + a^2)}{\Delta}\mathcal{L}_F; \quad \epsilon_\theta = \mathcal{L}_F \quad (38)$$

The magnetic induction $\mathbf{B}$ is related with the magnetic field intensity $\mathbf{H}$ by

$$B_\alpha = \mu_{\alpha\beta} H_\beta \quad (39)$$

where $\mu_{\alpha\beta}$ is the tensor of the magnetic permeability whose independent eigenvalues are

$$\mu_r = \frac{(r^2 + a^2)}{\Delta}\frac{1}{\mathcal{L}_F}; \quad \mu_\theta = \frac{1}{\mathcal{L}_F} \quad (40)$$

In the de Sitter region $\epsilon_r = \epsilon_\theta = \mathcal{L}_F$; $\mu_r = \mu_\theta = \mathcal{L}_F^{-1}$.

The dynamic equations (32) yield

$$\mathcal{L}_F \Sigma^2 F_{10} = e(r^2 - a^2\cos^2\theta); \quad \mathcal{L}_F \Sigma^2 F_{20} = -era^2\sin 2\theta \quad (41)$$

and the components $F_{13}$ and $F_{23}$ are given by (34).

The stress-energy tensor of a nonlinear electromagnetic field is calculated in the standard way [34] from the electromagnetic lagrangian $\mathcal{L}(F)$ which gives

$$\kappa T_\nu^\mu = 2\mathcal{L}_F F_{\nu\alpha} F^{\mu\alpha} - \frac{1}{2}\delta_\nu^\mu \mathcal{L} \quad (42)$$

The equation of state in the co-rotating frame

$$\kappa(p_\perp + \rho) = 2\left(\mathcal{L}_F F_{10}^2 + \mathcal{L}_F \frac{F_{20}^2}{a^2\sin^2\theta}\right) \quad (43)$$

allows one to investigate the behavior of the fields on the de Sitter vacuum disk (30).

Expressing the field components through their values given by (41), we get two equations governing the field dynamics of a spinning electrovacuum soliton:

$$\kappa(p_\perp + \rho) = \frac{2e^2}{\mathcal{L}_F \Sigma^2} \quad (44)$$

$$\mathcal{L}_F^2 F\Sigma^2 = -2e^2 + \frac{16e^2 r^2 a^2 \cos^2\theta}{\Sigma^2} \quad (45)$$

The equations (43)-(45) are valid everywhere.

In the second term of eq. (45) we have

$$\frac{r^2 a^2 \cos^2\theta}{\Sigma^2} = \frac{r^2}{\Sigma} - \frac{r^4}{\Sigma^2} \quad (46)$$

In the equatorial plane it vanishes by (28), as a result

$$\mathcal{L}_F^2 F\Sigma^2 = -2e^2 \quad (47)$$

Combining (47) valid in the equatorial plane with (44) valid everywhere, we get in the equatorial plane

$$\mathcal{L}_F F = -\kappa(p_\perp + \rho) \quad (48)$$

Equations (47) and (48) give in the equatorial plane

$$\mathcal{L}_F = \frac{2e^2}{\Sigma^2 \kappa(p_\perp + \rho)}; \quad F = -\frac{\kappa^2(p_\perp + \rho)^2 \Sigma^2}{2e^2} \quad (49)$$

Sign of $\mathcal{L}_F$ is the same as the sign of $(p_\perp + \rho)$ defined by (29). For original spherically symmetric soliton satisfying the weak energy condition, $\tilde\rho' \leq 0$ everywhere [33].

On the equatorial disk the left hand side of (43) goes to zero by (30), which means that each component in the right hand side vanishes on the disk independently

$$\mathcal{L}_F \frac{F_{20}^2}{a^2\sin^2\theta} = 0; \quad \mathcal{L}_F F_{10}^2 = 0 \quad (50)$$

The field invariant $F$ goes to zero by (49).

On the disk the dielectric permeability $\epsilon_r = \epsilon_\theta = \mathcal{L}_F$ goes to infinity by (49), and the magnetic permeability $\mu_r = \mu_\theta = \mathcal{L}_F^{-1}$ goes to zero, so that the rotating de Sitter vacuum disk displays both perfect conductor and ideal diamagnetic behavior.

The magnetic induction **B** goes to zero on the disk independently on the magnetic permeability. Indeed, it follows from (50) that on the disk

$$\frac{2e^2(B^r)^2}{\kappa(p_\perp+\rho)(r^2+a^2)^2} = 0; \quad \frac{2e^2(B^\theta)^2}{\kappa(p_\perp+\rho)a^2\sin^2\theta} = 0 \quad (51)$$

so that $(B^r)^2$ and $(B^\theta)^2$ must vanish faster than $(p_\perp+\rho)$. The Meissner effect takes place for a single spinning soliton and occurs at its de Sitter vacuum disk.

On the intrinsically flat disk we can apply the conventional definition of the surface current [35]

$$\mathbf{g} = \frac{(1-\mu)}{4\pi\mu}[\mathbf{nB}] \quad (52)$$

where **n** is the normal to the surface.

The condition (52) relates the current density **g** to the magnetic induction inside a body and thus to the currents at every point on the surface. The transition to a superconducting state corresponds formally to the limit $\mathbf{B} \to 0$ and $\mu \to 0$. The right hand side of (52) then becomes indeterminate, and there is no condition which would restrict the possible values of the current [35]. As a result the surface currents on the de Sitter disk can be any and amount to a non-zero total value.

All this allows one to conclude that NED coupled to gravity suggests that spinning particles dominated by the electromagnetic interaction, would have to have de Sitter interiors arising naturally in the regular geometry asymptotically Kerr-Newman for a distant observer. De Sitter vacuum supplies a particle with the finite positive electromagnetic mass related to breaking of space-time symmetry. In this economic picture the interior rotating de Sitter vacuum disk displays kind of an elementary superconductivity within a single spinning particle.

This behavior revealed for an arbitrary nonlinear lagrangian $\mathcal{L}(F)$, is generic for the class of regular spinning electrovacuum solutions (for both black hole and particle) specified ultimately by a regular density profile $\tilde{\rho}(r)$ in (17). The explicit example of a spherically symmetric electrovacuum soliton is presented in [33], it satisfies the dominant energy condition which facilitates study (after transforming it to (10)) of scattering of photons by a spinning electrovacuum soliton.

NED theories appear as low-energy effective limits in certain models of string/M-theories (for review [36,37]). The above results apply to the cases when the relevant NED scale (for example the characteristic scale for leptons) is much less than the Planck scale.

**Acknowledgement -** This work was supported by the Polish Ministry of Science and Information Society Technologies through the grant 1P03D.023.27. I am very grateful to Jürgen Ulbricht and André Rubbia for encouragement at the early stage of this research.